\documentclass[%
reprint,
superscriptaddress,
showpacs,showkeys,
floatfix,
]{revtex4-1}

\usepackage{graphicx}
\graphicspath{{Figures/}}
\usepackage[normalem]{ulem}
\usepackage{bm}
\usepackage[T1]{fontenc}
\usepackage[table]{xcolor}
\usepackage[colorlinks=true,citecolor=blue,linkcolor=blue,urlcolor=blue]{hyperref}

\makeatletter

    \def\CT@@do@color{%
      \global\let\CT@do@color\relax
            \@tempdima\wd\z@
            \advance\@tempdima\@tempdimb
            \advance\@tempdima\@tempdimc
    \advance\@tempdimb\tabcolsep
    \advance\@tempdimc\tabcolsep
    \advance\@tempdima2\tabcolsep
            \kern-\@tempdimb
            \leaders\vrule
                    \hskip\@tempdima\@plus  1fill
            \kern-\@tempdimc
            \hskip-\wd\z@ \@plus -1fill }
    \makeatother

\begin{document}

\preprint{APS/123-QED}

\title{Tailoring interface alloying and magnetic properties in (111) Permalloy/Pt multilayers}

\author{Movaffaq Kateb}
\affiliation{Science Institute, University of Iceland, Dunhaga 3, IS-107 Reykjavik, Iceland}
\affiliation{Department of Engineering, School of Technology, Reykjavik University, Menntavegi 1, IS-102 Reykjavik, Iceland}
\author{Jon Tomas Gudmundsson}
\affiliation{Science Institute, University of Iceland, Dunhaga 3, IS-107 Reykjavik, Iceland}
\affiliation{Space and Plasma Physics, School of Electrical Engineering and Computer Science, KTH--Royal Institute of Technology, SE-100 44, Stockholm, Sweden}
\author{Snorri Ingvarsson}%
 \email{sthi@hi.is}
\affiliation{Science Institute, University of Iceland, Dunhaga 3, IS-107 Reykjavik, Iceland}

\date{\today}

\begin{abstract}
We demonstrate preparation and characterization of permalloy Ni$_{80}$Fe$_{20}$ at.~\% (Py) multilayers  in which the Py layers were deposited by two different sputter deposition methods. Namely, dc magnetron sputtering (dcMS) and high power impulse magnetron sputtering (HiPIMS), that represent low and moderate ionized flux fraction of the film forming material, respectively. We deposited ultrathin bilayers 15~{\AA} thick  Py and 5~{\AA} thick Pt with 20 repetitions. Various effects such as substrate roughness, working gas pressure and sputter power are considered. The microstructure, texture and strain were characterized by X-ray diffraction, individual thicknesses and alloying were analysed by X-ray reflectivity, and uniaxial in-plane anisotropy probed by the magneto-optical Kerr effect. It is shown that HiPIMS deposition produces multilayers with reduced surface roughness regardless of the substrate surface roughness. Both dcMS and HiPIMS deposition present multilayers with strong (111) texture normal to the substrate. Using HiPIMS for deposition of the Py layer minimizes the alloying between individual layers compared to dcMS deposition performed at same average sputter power. However, this improvement in interface sharpness leads to a higher magnetic coercivity and a poor hard axis in the film plane. On the other hand, multilayers with alloying present a linear hard axis. Furthermore, we studied Py/Cu and Py/CuPt multilayers, prepared under identical conditions using HiPIMS. The results indicate that in the Py/Pt case, deterioration of the in-plane uniaxial anisotropy, is caused by inverse magnetostriction originating from the large lattice mismatch between Py and Pt. The Py/Pt multilayers that exhibit alloying have a less strained interface and have a well defined uniaxial anisotropy.
\end{abstract}

\maketitle


\section{Introduction}
Magnetic multilayers consisting of ferromagnetic (FM)/non-magnetic (NM) layers have been widely studied for giant magnetoresistance (GMR) \citep{chen2020,torrejon2020,fathoni2019}, antiferromagnetic exchange coupling \citep{greaves2018}, ferromagnetic resonance (FMR) response, and magnetic damping \citep{wei2019,du2020}. Often, low magnetic damping multilayer stacks such as FM/Cu are preferred for GMR applications \citep{fathoni2019}. In some cases enhanced Gilbert damping is desirable, such as present in FM/Pt stacks, due to high spin-orbit scattering at the interfaces \citep{lazarski2019}. With carefully chosen Pt thickness these layers also exhibit enhanced anisotropic magnetoresistance (AMR), attributed to spin-orbit scattering and retention of magnetic order at the interfaces (suppression of a ``magnetic dead layer") \citep{liu2010}. These properties make Pt an unique candidate for more recent applications such as the FMR bandwidth modulation \citep{hayashi2021}, stabilization of skyrmions against thermal fluctuations \citep{legrand2020}, as well as femtosecond (fs) laser pulse induced photocurrent, and THz radiation \citep{medapalli2020}. A promising approach that can accelerate the recent applications is vertical packing of multiple stacks. Recently, this approach has been utilized for noise reduction within both GMR sensors and recording media \citep{torrejon2020,greaves2018}. However, it requires revisiting the preparation methods to achieve reduced thicknesses and tailored interface.

An important property of a FM/NM stack is its magnetic anisotropy \citep{johnson1996}. It has been shown that a decrease in the thickness of a single FM layer favours perpendicular anisotropy \citep{purcell1992,johnson1996}. This is even true for Permalloy Ni$_{80}$Fe$_{20}$ at.~\% (Py) \citep{johnson1996} that is known to present in-plane magnetic anisotropy when in the thin film form \citep{kateb2017,kateb2018,kateb2019,kateb2019epi,kateb2019phd,kateb21:167982,liu2010,hirayama2017}. In a multilayer stack, as each FM layer becomes thinner the effect of the interface on anisotropy can outweigh the bulk contribution of individual layers. As a result, the strong demagnetizing field which is responsible for forcing magnetization into the film plane can be overcome \citep{johnson1996}. This, observation motivated numerous studies on the crossover between in-plane and perpendicular anisotropy \citep{Monso:2002js,Dieny:2017ey}. The relation between interface alloying, or intermixing, with the magnetic anisotropy has, on the other hand, not received much attention \citep{den1988}. However, the impact of the interface on the magnetic damping \citep{ingvarsson2002}, interlayer exchange coupling \citep{azzawi2017} and THz emission \citep{torosyan2018,scheuer2018} has been widely studied. Reports on the influence of the alloying on e.g.\ the magnetic damping still appear to be controversial \citep{ganguly2015,azzawi2016,dominguez2015}. This is due to fact that there are two main approaches towards studying the interface effect. In the more common approach, the interface alloying is considered inherent to the deposition method \citep{johnson1996,holmstrom2004} and, instead of changing interface alloying at a constant thickness, it is assumed that increasing the FM thickness reduces the interface contribution. The other approach relies on post-deposition treatments, such as annealing \citep{den1988} or ion-beam irradiation \citep{ganguly2015}, for modification of the interface alloying. Often however, the change by such methods is not limited to the interface but it also applies to the interior of the individual layers \citep{woods2002,rafaja2002b}. Thus, it is important to have a technique that allows us to tailor the interface quality but leaving the film interior intact.

The minimum requirement of a method for multilayer preparation is continuous layers without droplet formation and bridging between layers \citep{shen2004}. For instance, using dc magnetron sputtering (dcMS) a minimum thickness of 18 and 39~{\AA} has been reported by our group for the coalescence of Pt islands and a continuous Pt film, respectively \citep{agustsson08:082006,agustsson08:7356}. The latter was achieved on a Si substrate and determined by \textit{in-situ} resistivity measurements during the growth, while the in-plane grain size and the surface roughness were determined by scanning tunneling microscopy (STM). Continuous Pt films of lower thicknesses can be achieved by increasing the
substrate temperature during deposition. Also, having Py as an underlayer, the minimum thickness for a Pt film to be continuous can be lowered, e.g.\ 9.4~{\AA} has been reported for a bilayer system \citep{azzawi2016}. In Py/Pt \emph{ multilayers}, continuous Pt layers down to 15 -- 20~{\AA} have been reported \citep{dominguez2015,correa2016}. It is generally believed that as the number of periods increases the interface roughness increases \citep{rafaja2002}, and consequently that limits the minimum Pt thickness. One may think this is probably the reason for limited studies on the Py/Pt multilayers, while its bi- and trilayers have been widely studied. However, it appears that even in the bilayers alloying of the NM with the FM is a major issue e.g.\ mixed interface of 9.2 -- 10~{\AA} is reported for Py/Pt \citep{azzawi2016,ganguly2015}, while another study reported intermixing up to 80~{\AA} for Pt/Py/Pt trilayers \citep{dominguez2015}.  Such a difference in the thickness of the mixed interface is an example of a debate that can rise around this topic that might be initiated by the deposition method. Another example is the controversial results on the existence of interlayer exchange coupling through the Pt spacer \citep{parkin1991b,lim2013,omelchenko2018}. However, \citet{parkin1991b} did not rule out problems caused by the deposition method and recently \citet{omelchenko2018} showed that previous studies \citep{lim2013} were limited to the Pt thickness above 20~{\AA} where the exchange coupling is too weak to be detected.

Many multilayer studies have been devoted to comparison of different deposition methods, often resulting in very different magnetic properties. For instance, it has been shown that pulsed laser deposition (PLD) gives improved layer-by-layer growth and (111)\,Fe/Cu, compared with molecular beam epitaxy (MBE) of the same layers. In fact, the magnetic anisotropy result in these cases is entirely different, i.e.\ the former has perpendicular anisotropy while the latter presents in-plane anisotropy \citep{shen2004,willmott00:315}. It is worth mentioning that a typical instantaneous deposition rate of PLD is 5 -- 6 orders of magnitude higher than that of MBE. The layer-by-layer growth at such a high deposition rate can be explained by the higher ionized flux fraction in the case of PLD \cite{kateb2019md,kateb2020md}.

High power impulse magnetron sputtering (HiPIMS) is another ionized physical vapor deposition technique \cite{helmersson06:1,gudmundsson12:030801}. In dcMS the film forming material consists mainly of neutral atoms while in HiPIMS deposition it consists of both ions and neutrals \citep{gudmundsson20:113001}. However the increased ionized flux fraction in HiPIMS operation comes at the cost of decreased deposition rate \citep{brenning20:033008}. Therefore, the average deposition rate of a typical HiPIMS deposition is lower than that of dcMS deposition \citep{samuelsson10:591}, while the instantaneous deposition rate during the pulse is high. In the dcMS operation utilizing high power density is limited by thermal load on the target, and a high deposition rate does not allow us to deposit ultrathin multilayers. The impulse nature of HiPIMS solves both of these issues i.e.\ during the pulse we have both high power  density and high deposition rate but the average power and average deposition rate can be modulated by adjusting the pulse length and repetition frequency \cite{helmersson06:1,gudmundsson12:030801}.  
The energetic adatoms in high power operation provide enough energy for texture refinement. The major advantages of HiPIMS deposition are known to be denser \cite{samuelsson10:591,magnus11:1621,magnus12:1045}, void-free \cite{alami05:278} and smoother thin films \cite{alami05:278,magnus11:1621} compared to dcMS deposited films. It is also worth noting that using molecular dynamics simulation we have demonstrated that highly ionized flux of the film forming material gives a superior smoothness compared to less ionized alternatives \citep{kateb2019md,kateb2020md}. Earlier, we have demonstrated that for a monolayer of Py HiPIMS deposition gives denser films with lower coercivity and anisotropy field than dcMS deposition \citep{kateb2018,kateb2019epi}. However, so far there is no report on utilizing HiPIMS for FM/NM multilayer preparation. This may be due to the presence of highly energetic ions that are \emph{expected} to increase intermixing between different layers. The latter may also be used as an argument against the use of PLD for multilayer deposition, but experimental studies have proven otherwise \citep{shen2004}.

In the present study, we investigate the effect of intermixing in the [Py (15~{\AA})/Pt (5~{\AA})]$_{20}$/Ta (50~{\AA})/SiO$_2$/Si multilayer upon variation of deposition parameters.  Py has been chosen because it meets several conditions: First, Py is a fcc metal that eliminates many complications that can arise e.g.\ in hcp Co. It has a relatively small lattice parameter and thus with the choice of NM layer it can be strained in a large window. Also, it has smaller saturation magnetization compared to Ni, Co and Fe that allows strain induced anisotropy to overcome shape anisotropy and small magnetocrystalline anisotropy. Finally, we have studied different aspects of Py thin films and are quite familiar with its growth and properties \citep{ingvarsson2017,kateb2017,kateb2018,kateb2019,kateb2019epi,kateb2019phd,kateb21:167982}. Section II discusses the methods and experimental apparatus, and Section III discusses the properties of the multilayers deposited with the various combinations of deposition techniques, applied power and substrates.  We address, strain and texture, interface roughness and alloying, and magnetic properties. A summary is given in Section IV.

\section{Experimental apparatus and method}

We used square 10$\times$10~mm$^2$ (001) p-Si  substrates with a $\sim$2.4~nm native oxide and 100~nm thermally grown oxide. The multilayers were deposited in a ultra-high vacuum co-deposition (multi-target) chamber that reaches a base pressure below $5\times 10^{-7}$~Pa. All samples were prepared at room temperature (21$\pm$0.1~$^\circ$C). During sputter deposition high purity argon (99.999\%) was used as the working gas. We utilized Ni$_{80}$Fe$_{20}$~at.\% (99.5\%) and  Pt (99.99\%) circular targets of diameters 75 and 50~mm, respectively. The substrate center was located 22~cm from the center of all targets. In all cases, prior to the deposition of the multilayers, we deposited a 50~{\AA} Ta as under-layer at 10~W using dcMS from a 50~mm diameter Ta target. The latter encourages $\langle111\rangle$ texture perpendicular to the substrate surface in the Py multilayers as shown by Ritchie \textit{et al.} \citep{ritchie2002}.

During the deposition, we rotated the sample holder 360$^\circ$ back and forth at $\sim$12.5~rpm to ensure thickness uniformity of the deposited films \citep{kateb2019}. 
This is necessary because in a co-deposition chamber the deposition flux arrives under an angle with respect to the substrate normal. Our chamber has a built-in 35$^\circ$ angle which allows us to induce in-plane uniaxial anisotropy in the Py without applying an external field \cite{kateb2017,kateb2018,kateb2019,kateb2019epi}. We have shown that the so-called tilt deposition can win over applying magnetic field during growth in definition of magnetic axes \citep{kateb2017,kateb2019,kateb2019epi}. Further details on the geometry of our deposition chamber have been reported elsewhere \cite{kateb2017}.

For HiPIMS deposition, we utilized a dc power supply (ADL GS30) along with a pulser unit (SPIK1000A, Melec GmbH) operated in voltage control mode (uni-polar negative constant voltage pulses) through an analog interface. To monitor the discharge voltage and discharge current a combined voltage divider and current transformer unit (Melec GmbH) was used. During the HiPIMS deposition, the discharge current and voltage were recorded using a data acquisition board (National Instruments) with high temporal resolution controlled by a LabVIEW program. The latter also controls the pulse length and repetition frequency which were set to be 250~$\mu$s and 100~Hz, respectively. We maintained the time-averaged HiPIMS power around 153~W by applying voltage pulses with 472--490~V amplitude. The latter depends on the working gas pressure and increases with a decrease in working gas pressure. Since a high working gas pressure drastically increases the surface roughness in both dcMS and HiPIMS deposition \citep{kateb2018}, we are limited to relatively low working gas pressure when depositing ultrathin multilayers. On the other hand, there is a minimum pressure at which HiPIMS characteristics, such as delay time and discharge current onset, change drastically \citep{kateb2018}. In the available pressure window, we studied the effect of the pressure by comparing deposition at 0.25 and 0.40~Pa. 

The dcMS depositions of Py layers were performed at 50 and 150~W using a dc power supply (MDX 500, Advanced Energy). The 50~W dcMS deposition power was chosen because it gives a deposition rate of 0.75~{\AA}/s, similar to the average deposition rate of HiPIMS at time-averaged power of 150~W, and 150~W dcMS was selected to be similar to the HiPIMS average power. Note that dcMS deposition at 150~W  gives a deposition rate of 1.5~{\AA}/s, that is twice of that using HiPIMS with the same average power. In all cases we used dcMS for deposition of the Pt layers. To this end we utilized 20~W dc power which gives a deposition rate of 0.45~{\AA}/s.

X-ray diffraction (XRD) analysis was performed using a diffractometer (X'pert PRO PANalytical) utilizing a Cu source, with  the Cu K$_{\rm \alpha}$ peak, wavelength 0.15406 nm. The diffractometer was mounted with a mirror and a collimator (0.27$^\circ$) on the incident and diffracted side, respectively. For some samples we utilized a Ni filter to remove the Cu K$_\beta$ and W  L$_\alpha$ peaks. However, the filter has no effect on the main multilayer peaks. We also utilized a line focus that gives a beam width of $\sim$1~mm onto the sample. To determine the film thickness, density and surface roughness, low-angle X-ray reflectivity (XRR) was utilized. The XRR scans were made with an angular resolution of 0.005$^\circ$. The results were fitted based on the \citet{parratt54:359} formalism and distorted-wave Born approximation \citep{vineyard1982,sinha1988} as implemented in the commercial X'pert reflectivity program.

In-plane magnetic hystereses were measured using a custom-built magneto-optical Kerr effect (MOKE) looper with high sensitivity. After exploring different angles in the film plane, the direction of hard and easy axes were determined. We utilized a variable step size, i.e.\ 0.1~Oe around the switching along the easy axis, 0.5~Oe for both low fields along the easy axis and for the hard axis, and larger steps after saturation to high fields. 

\section{Results and discussion}
\subsection{Strain and texture}
\begin{figure}[hbt!]
    \centering
    \includegraphics[width=1\linewidth]{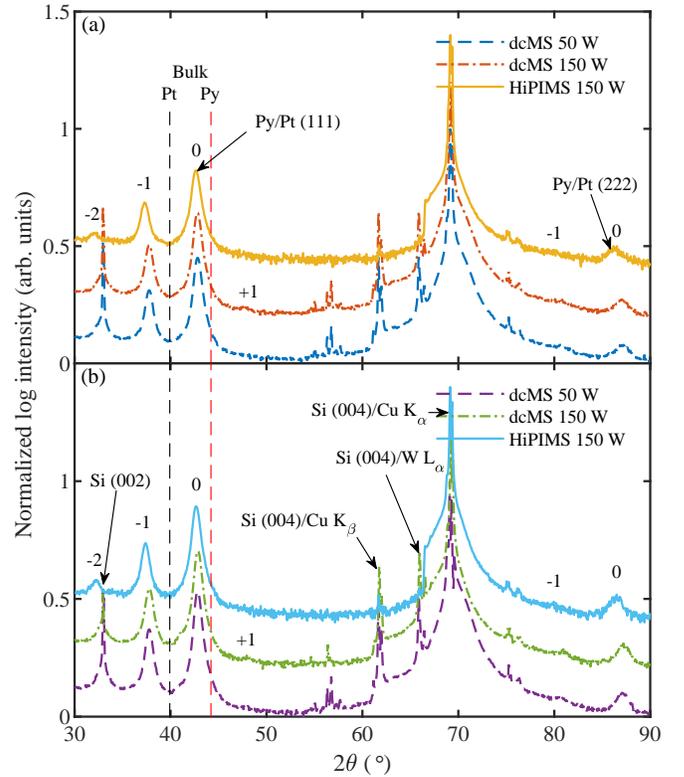}
    \caption{XRD patterns of Py/Pt multilayers deposited on p-Si (001) with (a) native oxide and (b) 100~nm thermally grown oxide. The legend indicates the sputtering method and power utilized for Py deposition. The vertical dashed lines in black and red indicate (111) peak position for the bulk Pt and Py, respectively. The XRD pattern for HiPIMS deposited multilayers were recorded using a Ni filter to remove the Cu K$_\beta$ and W L$_\alpha$ peaks which leaves a jump at 66.5$^\circ$.}
    \label{fig:XRDPower}
\end{figure}

Figure~\ref{fig:XRDPower} shows the XRD pattern from Py/Pt multilayers deposited on the p-Si (001) substrate with native oxide and 100~nm thermally grown oxide. The vertical dashed lines in black and red indicate (111) peak position for the bulk state Pt (ICDD 00-001-1190) and Py (ICDD 01-071-8324), respectively. It can be seen that a single (111) Py/Pt multilayer peak appears between the bulk Py and Pt peaks. Such a single multilayer peak have previously been reported for ultrathin Fe/V \citep{palsson2008}, Fe/Au \cite{rafaja2002b} and Co/Au \citep{den1988} (with 5.4, 12 and 13.1\% lattice mismatch, respectively). \citet{rafaja2002b} also showed that a hard annealing process can result in discontinuous layers that returns the XRD spectra to two separate peaks at the respective bulk peak positions \citep{rafaja2002b}. Thus, such a decomposition of a multilayer peak into peaks and their shift towards the bulk positions is an indication of discontinuous layers. For instance discontinuity as small as 15\% has been detected for Fe~(26~{\AA})/Au~(24~{\AA}) \citep{rafaja2002b}. For thinner layers, such as reported on here, smaller discontinuities should be detectable while thicker ones might already present a peak at the bulk position as strain has relaxed. Thus, our XRD results present characteristics of continuous layers. This indicates that all the Py and Pt layers are under in-plane tensile and compressive strain, respectively. This is expected due to the large 10\% lattice mismatch at the interface, that does not relax unless the thickness of the individual layers exceed a few~nm in thickness \citep{rafaja2002b,correa2016}. To the left of the (111) peak, there are two satellite peaks indicated by $-1$ and $-2$ and the distance to those is proportional to the multilayer period ($\Lambda$). It is worth mentioning that extracting period thickness from XRD at high angles might be tricky due to the fact that satellite peaks are caused by diffraction while XRR fringes are the result of interference as discussed by \citet{palsson2008}. Asymmetric intensity of satellite peaks, e.g.\ $+1$ \textit{versus} $-1$, is a characteristic of strain normal to the film plane \cite{vartanyants2000}. Here we do not observe positive satellite peaks except for a very small $+1$ peak for the multilayer deposited by dcMS at 150~W on both substrates and 50~W dcMS on the 100~nm thermally grown oxide substrate (These are hardly visible on the scale used in Figure~\ref{fig:XRDPower} but are magnified in insets in Figure~\ref{fig:MOKEPower}). Absence of the $+1$ peak is indicative of huge strain normal to the film caused by in-plane epitaxial strains. This strain has been shown to be relaxed by precise annealing, rendering the satellite peaks more symmetric \citep{den1988}. Thus, one can consider Py/Pt multilayers that exhibit $+1$ peaks to be slightly more relaxed than their counterparts. The sharp peaks in Figure~\ref{fig:XRDPower} are characteristics of single crystal substrate e.g.\ the largest peak at 69.9$^\circ$ is the Si (004) peak. The smaller sharp peak that sometimes appears at 34.6$^\circ$ belongs to Si (002) planes which is a structurally forbidden peak, although it sometimes shows a weak presence as discussed by \citet{zaumseil2015}. Due to the lack of Py/Pt (002) and (022) peaks, we conclude that these multilayers have a strong $\langle111\rangle$ texture normal to the substrate surface. Note that the Ta layer is found to be amorphous and thus it presents no peak in the XRD pattern \citep{ritchie2002}. It is also worth mentioning that the main (111) peaks observed from the multilayers prepared by HiPIMS deposition of the Py layers are slightly shifted towards smaller angles. Such a shift can be interpreted as a slightly larger lattice spacing normal to the film and a contraction in-plane. The latter can be considered as more pure Py layer, with less alloying with the Pt as will be discussed in more details in the following. 

\begin{figure}[hbt!]
    \centering
    \includegraphics[width=1\linewidth]{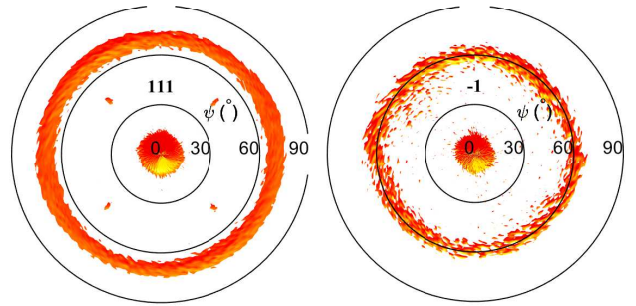}
    \caption{Pole figures of the main (111) Py/Pt multilayer peak (left) and its $-1$ satellite peak (right) obtained by dcMS deposition at 150~W on the 100~nm thermally grown oxide. The background is removed for better illustration.}
    \label{fig:pole}
\end{figure}

Furthermore, we studied our multilayers using polar mapping, i.e.\ scanning in-plane ($\phi$) and out-of-plane ($\psi$) angles while $\theta-2\theta$ angle is fixed to the main (111) multilayer or satellite peaks. As an example, the results for the main (111) and $-1$ satellite peaks for a multilayer deposited by dcMS at 150~W on the smooth 100~nm thermally grown oxide are shown in figure~\ref{fig:pole}. All the other (111) pole figures (not shown here) consisted of an intense spot at $\psi=0$, corresponding to the film normal, and a much weaker ring between $\psi=70-80^\circ$ corresponding to the (11$\bar{1}$) planes. Note that the relaxed (11$\bar{1}$) plane must appear at $\psi=70.5^\circ$. The dots at $\psi=45^\circ$ belong to the (113) Si planes. They are an indication of correct alignment of the instrument.

\subsection{Interface alloying}

\begin{figure}[hbt!]
    \centering
    \includegraphics[width=1\linewidth]{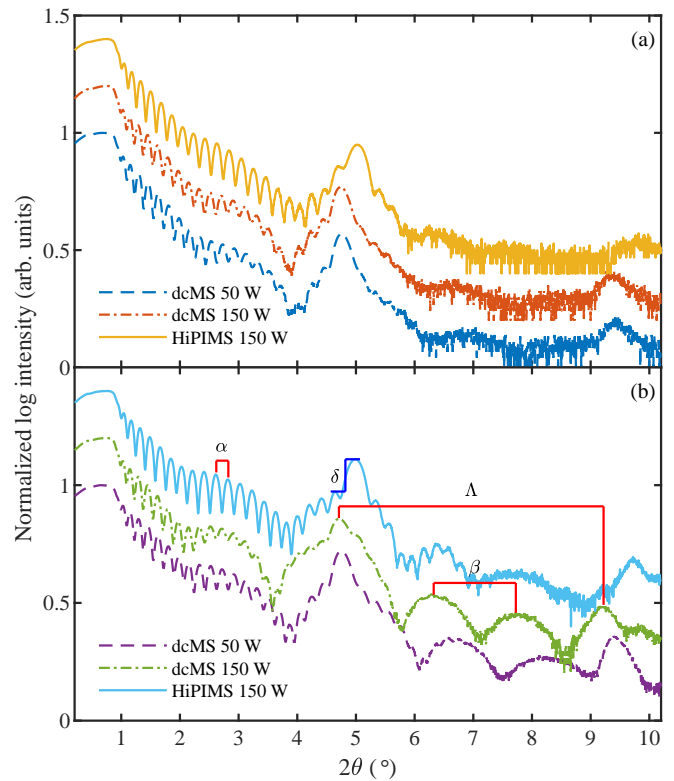}
    \caption{The result of XRR measurements for Py/Pt multilayers deposited on p-Si (001) with (a) native oxide and (b) 100~nm thermally grown oxide. The legend indicates the sputtering method and power utilized for Py deposition. $\alpha$, $\beta$ and $\Lambda$ are inversely proportional to the total thickness, Ta under-layer thickness and repetition thickness, respectively. The height $\delta$ is an indication of relative interface roughness with respect to the  overall roughness. The vertical dashed line indicates the location of the foremost Kiessig fringe.}
    \label{fig:XRRPower}
\end{figure}

Figure~\ref{fig:XRRPower} shows the XRR result of a Py/Pt multilayer deposited by dcMS and HiPIMS on (001) p-Si with native oxide and 100~nm thermally grown oxide. The values of $\alpha$, $\beta$ and $\Lambda$ indicated in the figure are inversely proportional to the total thickness, Ta under-layer thickness and period thicknesses, respectively. Here, the Ta acts as a buffer layer to decrease substrate roughness, while also promoting texture. It can be seen that the $\beta$ fringes from the Ta layer can be detected only when utilizing a smooth substrate (100~nm thermally grown oxide). However, for the native oxide substrate, these fringes disappear due to high roughness at the SiO$_2$/Ta interface. Using both substrates the $\Lambda$ fringes can be detected up to 10$^\circ$ which essentially means that the Ta layer can successfully improve surface roughness enough for growing delicate multilayers. For the case of HiPIMS deposition on the 100~nm thermally grown oxide, the amplitude of the $\alpha$ fringes decay much slower with the angle of incidence and they are visible up to $2\theta=7.3^\circ$. This means that HiPIMS deposition of the Py layers gives a smoother surface for the whole stack. As mentioned earlier, surface roughness commonly increases with the number of periods. It appears here, however, that HiPIMS deposition contributes to smoothening of the Py layers and consequently smoother interfaces, resulting in a smoother top surface for the multilayer stacks.

It has been shown that the discontinuity of individual layers can cause a shift of the foremost Kiessig fringe, indicated by vertical dashed line, towards smaller angles \citep{rafaja2002}. We observed a negligible shift (0.005$^\circ$) to the left for dcMS deposition case at 50~W on the 100~nm thermally grown oxide substrate. On the other hand, multilayers deposited by HiPIMS on different substrates and pressures present 0.01 -- 0.02$^\circ$ shift towards higher angles. Thus, based on this criteria there is lower probability of discontinuous layers within multilayers prepared by HiPIMS than by dcMS.

To determine alloying of the interfaces, we fit the XRR results carefully. The XRR fit gives the thickness and the density of the individual layers, and gives the interface roughness quantitatively. It has been shown that the XRR technique is a reliable method to study multilayers i.e.\ thickness and discontinuity of each layer as well as interface roughness \citep{den1988,chason1997,moreau2016thesis,azzawi2016}. It is worth mentioning that the interface roughness ($\delta$) is not an absolute value and changes in relation with the roughness of the whole stack \citep{baumbach1999} (cf.~$\delta$ in figure~\ref{fig:XRRPower} (b) that is proportional to $\Lambda$ and $\alpha$ intensity). For this reason, during the fitting process we gave more weight to the density and thickness and thus sacrificed the precision of the interface roughness. It is also worth mentioning that thanks to the large probing area of the XRR method (on the order of millimeters), surface roughness larger than film thickness is detectable \citep[\S~3.2.1]{foll2019}. We use the density of each layer to determine the interface mixing considering the fact that Pt layers shrink and Py layers expand in the film plane (cf.~XRD results). Thus, for an ideal Pt layer it is expected that its density increases above the bulk value of 21.45~g/cm$^3$. On the other hand, substitution of Pt with Ni and Fe that occurs at a mixed interface reduces the density due to much lower mass of the substituent. Using the same principle, we expect a density that is lower than 8.72~g/cm$^3$ for a pure Py layer due to lattice strain, while diffusion of Pt into the layer will counteract the decrease in density. As pointed out by \citet{baumbach1999}, for rough multilayers, such as the ones deposited on the native oxide, it is important to understand the interface morphology, e.g.\ stepped \textit{versus} Gaussian and its replication distance, before the fitting process. We apply the same fitting procedure to multilayers deposited on both substrates that may lead to inaccurate results for the native oxide.

The results of the fitting are summarized in table~\ref{XRRfit}. Utilizing dcMS deposition on native oxide gives higher density for the Py layer than when it is deposited  by the HiPIMS counterpart. The higher density indicates the existence of Pt in the Py layer which is evidence of a diffused interface. Although utilizing HiPIMS when depositing on native oxide can reduce the Pt diffusion, the value 8.98~g/cm$^3$ is still above 8.72~g/cm$^3$ (the bulk value), and the single layer density observed in our previous works (8.6 -- 8.7~g/cm$^3$ \citep{kateb2017,kateb2018}). The density of the Pt layers is higher than the bulk value of 21.45~g/cm$^3$ for the dcMS deposited cases as expected from the compressive strain. However, the Pt layer density is lower than the bulk value for the HiPIMS deposited counterparts which is evidence of Ni and Fe diffusion into Pt.
\renewcommand{\arraystretch}{1.2}
\begin{table}
  \caption{\label{XRRfit} Average values of Py and Pt thickness ($t$), density ($\rho$) and interface roughness ($R$)\footnote{Note that during the fitting process more weight is given to the density and thickness. This makes the resulting roughnesses less precise.} obtained by fitting the XRR measurement results.}
  \footnotesize{
  \begin{tabular}{c c c c | c c c c c c}
    \toprule
    Deposition & Power & substrate & $p$ & $t_{\rm Py}$ & $t_{\rm Pt}$ & $\rho_{\rm Py}$ & $\rho_{\rm Pt}$ & $R_{\rm Py}$ & $R_{\rm Pt}$ \\
    method & (W) & type & (Pa) & \multicolumn{2}{c}{({\AA})} & \multicolumn{2}{c}{(g/cm$^3$)} & \multicolumn{2}{c}{({\AA})} \\
    \hline
    \rowcolor{gray!40} dcMS & 50 & Native & 0.4 & 16.64 & 2.22 & 9.56 & 23.53 & 5.25 & 6.49 \\
    \rowcolor{gray!10} dcMS & 50 & 100~nm & 0.4 & 16.1 & 2.72 & 9.27 & 23.19 & 2.96 & 4.73 \\
    \rowcolor{gray!0} dcMS & 50 & 100~nm & 0.25 & 16.31 & 2.86 & 9.52 & 23.54 & 3.77 & 6.08 \\
    \hline
    \rowcolor{gray!40} dcMS & 150 & Native & 0.4 & 16.79 & 2.11 & 9.53 & 23.54 & 5.11 & 6.45 \\
    \rowcolor{gray!10} dcMS & 150 & 100~nm & 0.4 & 14.81 & 4 & 8.72 & 20.69 & 3.43 & 6.34 \\
    \rowcolor{gray!0} dcMS & 150 & 100~nm & 0.25 & 15.07 & 4 & 8.72 & 21.31 & 3.6 & 6.06 \\
    \hline
    \rowcolor{gray!40} HiPIMS & 150 & Native & 0.4 & 13.63 & 4.16 & 8.98 & 19.26 & 5.02 & 6.65 \\
    \rowcolor{gray!10} HiPIMS & 150 & 100~nm & 0.4 & 14.34 & 4.21 & 8.61 & 23.54 & 3.14 & 4.89 \\
    \rowcolor{gray!0} HiPIMS & 150 & 100~nm & 0.25 & 13.94 & 4 & 8.68 & 23.54 & 3.9 & 5.09 \\
    \botrule
  \end{tabular}
}
\end{table}

It was expected that using the smooth 100~nm thermally grown oxide substrate would effectively enhance the sharpness of the interfaces. However, for Py layers deposited by dcMS at 50~W, the densities indicate a negligible improvement compared to the native oxide, i.e.\ there is still diffusion of Pt into the Py layer. Depositing at 150~W with dcMS, the Py and Pt densities reach 8.72 and 20.69{ -- 21.31}~g/cm$^3$, respectively. Thus, using a smooth substrate, an increase in dcMS power reduces the diffusion of Pt into Py but it causes diffusion of Ni and Fe into Pt. On the other hand, HiPIMS deposition gives the lowest Py density of 8.61{ -- 8.68}~g/cm$^3$, indicating the least diffusion of Pt into Py, accompanied by a high density of 23.54~g/cm$^3$ for the Pt layers, indicating negligible diffusion of Ni and Fe into Pt.

For working gas pressure of 0.25~Pa, dcMS deposition at 50~W on a substrate with 100~nm thermally grown oxide gives higher density for both Pt and Py than a deposition at a pressure of 0.4~Pa. This means even more diffusion of Pt into Py but little diffusion of Ni and Fe into Pt. For 150~W dcMS deposition the Py density does not change with the working gas pressure but the Pt density increases, i.e.\ there is less diffusion of Ni and Fe compared to at 0.4~Pa. We also observed that there is a noticeable change in $\Lambda$ ($t_{\rm Py}+t_{\rm Pt}$) with pressure changes for 50~W dcMS. This means that at low power dcMS operation the deposition rate, and consequently the thickness of individual layers, are very sensitive to the variation in the working gas pressure. Thus, depositing with dcMS at low power, provides an extremely narrow and inflexible pressure window, which is undesirable. The latter can be improved by depositing at 150~W dcMS, i.e.\ a limited variation in $\Lambda$ in the 0.25 -- 0.4~Pa pressure range. Using HiPIMS, there is a negligible change in the Py density and the Pt density is maintained at the lower pressure. However, the multilayer period becomes smaller, which indicates that lowering the pressure decreases the deposition rate of Py. This is probably due to the delay between voltage and current onset which shows larger change with pressure variation below 0.33~Pa \citep{kateb2018}.

A question that may arise here is how an increase in sputter power by going from 50 to 150~W dcMS, and a further increase in instantaneous power by going to HiPIMS deposition decreases the interface alloying. Regarding the alloying by energetic ions, it has been shown that ions with bombarding energy of a few hundred eV can only penetrate into a few nm depth if the mass density is lower than 80\% of theoretical mass density \citep{muller1986,muller1986jap}. Using HiPIMS for the Py deposition leads to a higher film mass density which makes such penetration rather impossible. However, thermal spikes associated with energetic collisions might in-fact redistribute atoms at the interface \citep{kateb2019md}. This can be considered as inherent alloying. Thus, higher power here does not contribute to interface alloying in the conventional way of diminishing of the interface through bombardment and displacement of sublayer atoms. Quite the opposite, at higher powers the top surface becomes smoother and there is no increase in roughness with increased number of periods.
We have recently shown, using an atomistic simulation, that the increase in the ionized flux fraction in fact reduces surface roughness of a single metallic layer \citep{kateb2019md,kateb2020md}. The lower roughness is evident from the extent of $\alpha$ fringes at higher incident angles in Figure~\ref{fig:XRRPower} (b). A similar conclusion has been made by comparison of PLD with MBE. Although MBE grown layers are atomically sharp (1~{\AA}) in terms of alloying, the interface roughness might become tens of nm \cite{johnson1996}. On the other hand highly ionized deposition methods such as PLD and HiPIMS provide smooth surfaces and their interface alloying can be controlled e.g.\ by the correct choice of pulse length or other deposition parameters. 

\subsection{Magnetic properties}
\begin{figure}[hbt!]
    \centering
    \includegraphics[width=1\linewidth]{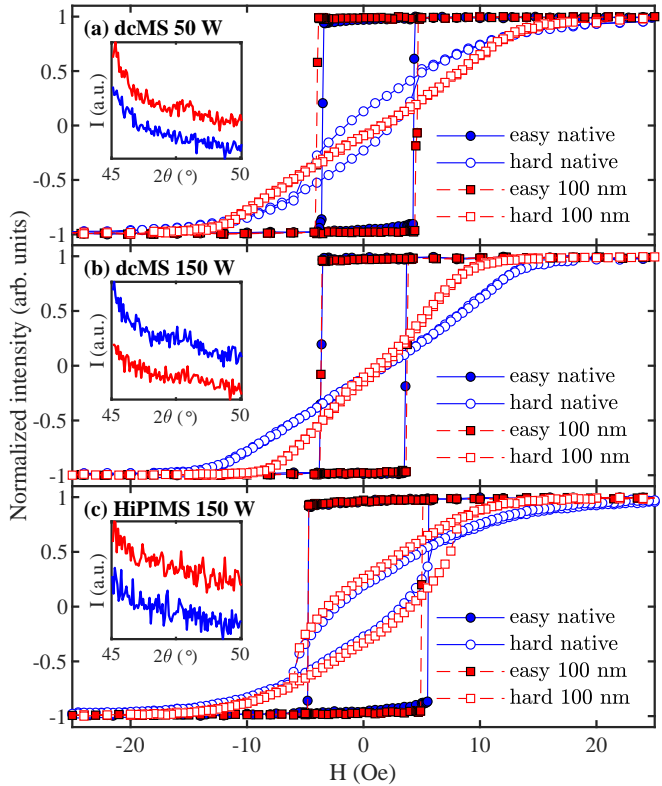}
    \caption{The MOKE response of Py/Pt multilayers deposited by (a) 50~W dcMS, (b) 150~W dcMS and (c) 150~W HiPIMS at 0.4~Pa. The probing direction is indicated by filled and hollow symbols while circle and square indicate native and 100~nm thermally grown oxides, respectively. The figure insets show a portion of XRD pattern (figure~\ref{fig:XRDPower}) where $+1$ satellite peak is expected. The blue and red curves belong the native and 100~nm thermally grown oxides, respectively.}
    \label{fig:MOKEPower}
\end{figure}

To address the question of how switching properties and in-plane anisotropy is affected by intermixing we measured the magnetic hysteresis loops by MOKE.  The MOKE experiments do not yield absolute values of magnetization, so the traces have been normalized to the saturation value in each case. Prior to these measurements, the direction of the magnetic easy axes had been determined by measuring hysteresis loops with the applied field at different angles in the film plane. The legend in each subfigure indicates the axis along which magnetic field is applied, and substrate type.  For the 50~W dcMS deposited stack shown in figure~\ref{fig:MOKEPower} (a), the substrate with native oxide gives an open hard axis trace and a well defined easy axis, while utilizing a smooth 100~nm thermally grown oxide substrate, enhances the squareness of the easy axis and presents a perfectly closed and almost linear hard axis trace. For the 150~W dcMS deposited stack, as shown in figure~\ref{fig:MOKEPower} (b), both substrates result in very well defined hard and easy axes. In this case, the 100~nm thermally grown oxide substrate results in a smaller anisotropy field, $H_{\rm k}$, but it also leads to a small hysteresis in the middle of the hard axis. This is barely visible in the figure, but is distinguished by a slope change in the hard axis loop at a around $\pm$4~Oe. 
%

Figure~\ref{fig:MOKEPower} (c) shows the magnetic response of the HiPIMS deposited stack which exhibits an open hard axis and square easy axis regardless of the substrate type (different smoothness). Note that these loops were obtained after examining different angles in the film plane. Thus they present the hardest and easiest directions in the film plane. The coercivity in this case is slightly higher than for both the dcMS counterparts. In general, opening in hard axis loops can be caused by domains whose magnetization directions deviate slightly from the easy axis direction, for whatever reason. This could be brought on by grain structure, surface roughness, or by interfacial strain. It has been shown that alloying Py with Pt can cause an increase in coercivity but the reports on the anisotropy field are controversial \citep{chen1991,ingvarsson2004}. This is not the case here since we have shown by fitting the XRR results that there is minimum alloying for multilayers prepared by HiPIMS deposition compared to their dcMS counterparts on the same substrate. 
The insets in Figure~\ref{fig:MOKEPower} display a blown up portion of the XRD data in figure~\ref{fig:XRDPower}, where the $+1$ satellite peak is expected, between 47.5 -- 48$^\circ$. As stated above, strain relaxation is accompanied by the appearance of the $+1$ satellite peak, and more symmetric satellite peaks in general. It is interesting to note that the open hard axis loops occur when there is no discernible $+1$ satellite peak, i.e.\ in the most strained films, HiPIMS deposited Py (Figure~\ref{fig:MOKEPower} (c)), or when it is barely detectable as in dcMS deposited Py at 50~W on native oxide and dcMS deposited Py at 150~W on 100~nm thermal oxide. The samples in which there are hints of the $+1$ satellite peak all have perfectly closed hard axis loops. This suggests a magnetoelastic response to the high epitaxial strain at the sharp Py/Pt interfaces, i.e.\ inverse magnetostriction, is responsible for the opening in the hard axis loops and increased coercivity. In dcMS deposited counterparts, such a strain is relaxed by e.g.\ diffusion of Pt with larger atomic radius into stretched Py or \textit{vice versa}. Except for the case of 50~W dcMS deposition  on native oxide, which similar to HiPIMS does not present $+1$ satellite peak (cf.~Figure~\ref{fig:XRDPower}), and is more strained. 
A summary of the anisotropy field ($H_{\rm k}$) and coercivity ($H_{\rm c}$) of multilayers prepared by both methods and different pressure can be found in table~\ref{tab:MOKE}.
\begin{table}[hbt!]
    \caption{Summary of anisotropy field ($H_{\rm k}$) and coercivity ($H_{\rm c}$) for both dcMS and HiPIMS deposited multilayers determined by MOKE.}
    \label{tab:MOKE}
    \centering
    \footnotesize{
    \begin{tabular}{c c c c | c c}
    \toprule
    Deposition & Power & substrate & $p$ & $H_{\rm k}$ & $H_{\rm c}$ \\
    method & (W) & type & (Pa) & \multicolumn{2}{c}{(Oe)} \\ \hline
    \rowcolor{gray!40} dcMS & 50 & Native & 0.4 & 13.5 & 3.9 \\
    \rowcolor{gray!10} dcMS & 50 & 100~nm & 0.4 & 13.5 & 4.3 \\
    \rowcolor{gray!0} dcMS & 50 & 100~nm & 0.25 & 13.5 & 5.3 \\ \hline
    \rowcolor{gray!40} dcMS & 150 & Native & 0.4 & 13.5 & 3.6 \\
    \rowcolor{gray!10} dcMS & 150 & 100~nm & 0.4 & 8.5 & 3.7 \\
    \rowcolor{gray!0} dcMS & 150 & 100~nm & 0.25 & 13.7 & 4.7 \\ \hline
    \rowcolor{gray!40} HiPIMS & 150 & Native & 0.4 & 10.7 & 5.2 \\
    \rowcolor{gray!10} HiPIMS & 150 & 100~nm & 0.4 & 9.5 & 4.9 \\
    \rowcolor{gray!0} HiPIMS & 150 & 100~nm & 0.25 & 14.7 & 4.8 \\
    \botrule
    \end{tabular}
    }
\end{table}

\begin{figure}[hbt!]
    \centering
    \includegraphics[width=1\linewidth]{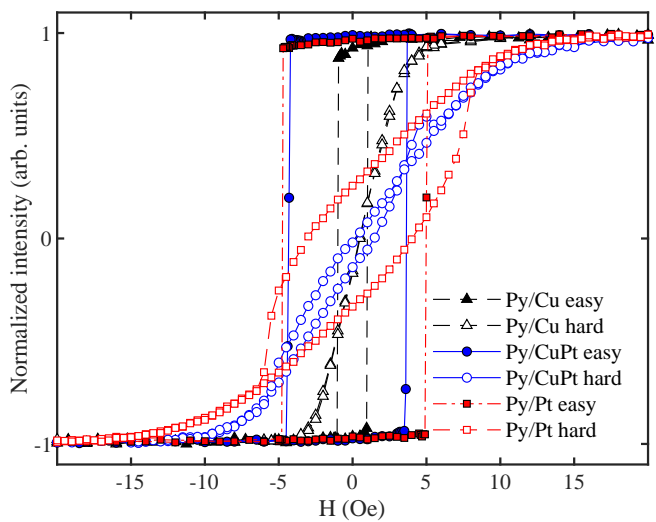}
    \caption{The MOKE response of Py/Cu, Py/CuPt and Py/Pt multilayers deposited by HiPIMS at 0.25~Pa on p-Si (001) with a smooth 100~nm thermally grown oxide. The legend indicates multilayers and the probing direction.}
    \label{fig:MOKEPyM}
\end{figure}

Furthermore, we studied the effect of strain by substituting Pt with Cu and Cu$_{50}$Pt$_{50}$~at.~\%. Cu has a lattice constant of 3.61~{\AA} close to that of Py (3.54~{\AA}) while Pt has a much larger lattice constant of 3.92~{\AA}. Figure~\ref{fig:MOKEPyM} shows the MOKE response of Py/Cu, Py/CuPt, and Py/Pt, multilayers deposited on the substrate with 100~nm thermally grown oxide. All the samples were deposited using identical conditions i.e.\ using HiPIMS for Py deposition and dcMS for NM layer deposition at a constant deposition rate. It can be seen that for the Py/Cu multilayers, very soft, but also very well defined uniaxial anisotropy is obtained. However, as the lattice constant of the NM increases upon substitution of Cu by Pt, the increase in strain at the interface leads to a higher coercivity and opening up in the hard axis trace. This is in agreement with the results for the single layer Py \citep{ohtani2013,kim1996,hollingworth2003,hollingworth2003b}. It has been shown that both polycrystalline (but textured) and single crystal (111) Py present almost isotropic magnetization in the film plane due to the strain \citep{ohtani2013}. It is worth mentioning that there exists a report on the correlation between magnetostriction coefficient and surface roughness \citep{kim1996}. In the latter study, the surface roughness changed as a consequence of change in layer thickness. Later, it was shown that surface roughness is rather unimportant compared to the effect of interfacial strain \citep{hollingworth2003,hollingworth2003b}. 

\section{Summary}

In summary, it is shown that ultrathin multilayers of Py~(15~{\AA})/Pt~(5~{\AA}) can be deposited successfully using HiPIMS. This means we are crossing the minimum dcMS limits i.e.\ (15~{\AA}) for continuous Pt layer and Py~(150~{\AA})/Pt~(120~{\AA}) for sharp interfaces. Multilayers obtained by both methods showed a strong (111) texture parallel to the substrate normal, supported by clear evidence of epitaxial strain. However, in comparison with dcMS, HiPIMS deposition provides a smoother surface and sharper interface, desirable traits when depositing multilayers. It has been shown that the dcMS technique is very sensitive to the working gas pressure and substrate roughness, and the former can be improved by an increase in the sputter power. Depositing with dcMS at low power, surface roughness and interface alloying and consequently magnetic properties depend on the substrate roughness. A higher dcMS power enhances uniaxial magnetic anisotropy and also interface sharpness. In the case of HiPIMS deposition, a sharp interface is obtained which indicates increased strain at the Py/Pt interfaces that leads to diminishing uniaxial magnetic anisotropy. Using Cu and CuPt instead of Pt it is verified that changing the interface strain can have a major influence on the resulting magnetic anisotropy of the multilayer.

\begin{acknowledgments}
This work was partially supported by the Icelandic Research Fund Grant Nos.~196141, 130029 and 120002023.
\end{acknowledgments}


\bibliography{Ref,heim78}
\end{document}